\documentclass[12pt]{article}
\usepackage{cite,amsmath,amsfonts,amsthm,fullpage}
\usepackage{epsfig}
\usepackage{eurosym}
\usepackage{verbatim}
\begin{document}
\baselineskip 16pt

\begin{center}
\begin{Large}\fontfamily{cmss}
\fontsize{17pt}{27pt} \selectfont \textbf{Approaches to Open Access in Scientific Publishing}
\end{Large}\\
\bigskip
\begin{large}  {J. Harnad}
\end{large}
\\
\bigskip
\begin{small}
 {\em Centre de recherches math\'ematiques,
Universit\'e de Montr\'eal\\ C.~P.~6128, succ. centre ville, Montr\'eal,
Qu\'ebec, Canada H3C 3J7} 
\end{small}
\end{center}
\bigskip

%%%%%%%%%%%%%%%%  Abstract  %%%%%%%%%%%%%%%%
\begin{center}{\bf Abstract}
\end{center}
\smallskip

\begin{small}
Approaches to scientific journal publishing that provide free access to all readers  are challenging the standard subscription-based model. But in domains that have a well-functioning system of publicly accessible preprint repositories like arXiv, Open Access is already effectively available. In physics, such repositories have long coexisted constructively with refereed, subscription based journals.   Trying to replace this by a system based on journals whose revenue is derived primarily from fees charged to authors  is unlikely to provide a better guarantee of  Open Access, and may be in conflict with the maintenance of high quality standards.
\end{small}
\medskip 

%%%%%%%%%%%%%%%%%  Section 1  %%%%%%%%%%%%%
\section{Open Access publishing: journals vs. repositories }

     The objective of Open Access (OA)  in scientific publishing is to make the results of research freely  available to all readers, thereby increasing both accessibility and dissemination.
However, from a publishers'  viewpoint,  making the contents of a journal available to all free of charge, even if only online,  implies loss of subscription revenue,  requiring that income be generated by other means, such as fees charged to authors.  For some non-profit publishers alternative sources  may be available,  such as professional association dues, or direct subsidies from host institutions, benevolent foundations or other funding organizations. But for commercial publishers, besides advertising revenue, the only real alternative to subscription fees is  transferring the cost burden from journal subscribers to authors, leading to an ``Author pays'' approach to OA. 

      In most domains of physics, however, and increasingly in mathematics and computer science, OA is already effectively implemented, at no cost to authors, through use of preprint repositories such as  arXiv. It seems difficult to argue the need for OA journals financed by publication charges to authors, just  to assure online availability of papers in a version formatted by the journal, when they are already freely available to all, at least in preprint form, via such repositories.

    There remain good reasons, of course, for publishing in refereed journals; in particular, the ``value-added" of quality control assured by the peer-review process. However, the cost of organizing this service,  the core of which is currently provided by the scientific community at no charge to the publisher, is far too small to justify the rates that some journals currently charge as subscription or publication fees.  Moreover, in the ``Author pays" approach to OA, rejection of submitted papers diminishes the number of articles published, and hence  decreases revenue to the publisher. Commercial logic tends therefore to militate against the application of overly high acceptance standards in this model, since it may be an impediment to publishing the quantity of articles needed to provide revenue adequate to sustaining the journal.
      
     Other factors, such as the  journal's  reputation, enter into consideration as well; authors prefer to publish in journals known for their high standards, and this provides a good motive for maintaining quality. But it is not clear, especially for a commercially based publisher, how best to balance these factors in order to ensure viability.   A possible implication may be that authors whose papers do not meet the peer review standards of the better subscription based journals, but are able and willing to pay  publication fees, could expect higher chances of acceptance in commercially run ``Author pays" OA journals. On the other hand, authors without the necessary funds, or unwilling to use scarce research funding for the purpose of paying publication fees, might prefer to publish in journals whose revenue does not depend on viewing authors as paying clients.

%%%%%%%%%%%%%%%%%  Section 2  %%%%%%%%%%%%%
\section{Open access journals, old and new }  
    
   The New Journal of Physics (NJP), published by the Deutsche Physikalische Gesellschaft (DPG) and the Institute of Physics (IOP), was a pioneer online OA journal that succeeded,  since 1998, in publishing a steady volume of articles that are widely read and cited. In fact, of the fifty-one journals listed under  ``physics" in Lund University's Directory of Open Access Journals (DOAJ) (only thirteen of which merit mention in the Scientific Citation Index), NJP is  the only one that is currently self-supporting without the need for institutional or government subsidies. But only recently has it been able to generate enough revenue from publication fees  to cover its production costs. Sustaining a journal that operates at a loss for many years is a choice that can be made by publishers answerable to non-profit professional associations like DPG and IOP, but not to  the  Board of Directors of a commercially motivated firm. 
      
       Nevertheless, with the appeal of  low production cost, rapid turnover online publishing, and with rising resistance within the user community to inflated subscription rates, several new online journals have been created by commercial publishers in recent years using the ``Author pays" model of OA. Biomed Central (BMC), for example, a UK-based  commercial publisher, founded in 2000 and acquired in 2008 by Springer-Verlag, has succeeded in running a large number of online OA journals in the biomedical field with revenues mainly generated by publication fees. Two new online physics journals were launched by  BMC in 2007 and 2008: PhysMath Central (PMC) Physics A (high energy physics) and PMC Physics B (condensed matter and atomic physics), with revenue generated by levying ``processing charges"   of  \euro{1100} per article. The number of papers appearing so far, however, suggests little enthusiasm on the part of authors from the physics community for paying such rates; in all, only four research articles have appeared in Physics A in 2008 and thirteen in Physics B.

   The announced policies of Biomed Central also provide some reason for  concern about the implementation of peer review standards.  In their ``instructions to editors", the editorial guidelines of several BMC journals have until recently stated: ``In the absence of  compelling reasons to reject, the Journal ... recommends acceptance, as ultimately the quality of an article will be judged by the scientific community after its publication." The editor of PMC Physics A has affirmed, however, that no such editorial guidelines are used by them and that high peer review standards are the norm. It is nevertheless difficult to see what would persuade researchers to pay a fee of  \euro{1100}  to have a refereed version of their paper made available for free online by the publisher, when the same work is already available to all via the arXiv, and can be published at no charge to the authors in a conventional subscription based journal.    
   
   Another publisher, Cairo-based  Hindawi Publishing Corporation,  founded in 1997, has a respectable track record in subscription-based scientific publishing. In 2007, it decided  to sell four of its standard subscription journals and convert another thirty to OA, launching a new range of about 150 OA journals in engineering,  medicine, physics, mathematics, biology and some other fields. A majority of these are still at an early start-up stage, posting  as yet no more than a handful of articles.   Only about half these journals have an editor-in-chief; the others are declared to be ``Community Journals", and have large editorial boards, but no scientific editor-in-chief to coordinate their operation. The combined number of editorial board members, garnered mainly through mass e-mail solicitations, is over five thousand. Although they are apparently trying to maintain adequate refereeing standards and  procedures, the absence of scientifically qualified coordinating editors in charge of their ``Community Journals" suggests a cutting of corners. They all rely strongly on automated methods of operation and communication, presumably with a view to minimizing production costs and keeping publication charges relatively low (between  \euro{200} and  \euro{400} per article for most of the new journals, but higher rates of up to \euro{1000} for the others).  Selection of referees and implementation of their recommendations is up to the board member chosen, but correspondence with referees is largely handled through an automated process of e-mail messages. These appear to be written, signed and sent by the board member, but the latter may, in fact, have never seen or approved the text.  It seems unlikely that many editorial board members will be happy to continue providing their services if such procedures are left unchanged, and suitably qualified editors-in-chief are not appointed for these journals, 
   
%%%%%%%%%%%%%%%%%  Section 3  %%%%%%%%%%%%%
\section{Community initiatives}

    Another significant movement is meanwhile appearing, one that undoubtedly provides cause for large commercial publishers to consider hedging their bets by acquiring low production cost, online journals based on the ``Author pays" model. This is the formation of large common interest groups within the scientific community, such as the SCOAP${}^3$ consortium in high energy physics, spearheaded by big laboratories like the CERN particle-physics laboratory, and now including institutional membership from eighteen countries.  SCOAP${}^3$ has the objective of encouraging a transition from subscription based journals to lower cost OA publishing through concerted action. With the addition of a sufficient number of  further  institutional partners,  this consumer alliance would represent a dominant share of research authors and journal readership in this field that could have enormous clout. 
        
   The size and strategy of the consortium has been evolving with time. After some rethinking of earlier approaches, such as transferring portions of their library budgets to subsidization of  individual journal publication charges for authors, they appear to be converging on the more effective strategy of offering the published work of their members on a competitive basis, as a package deal, and replacing the purchase of subscriptions or payment of individual publication fees by a global fee adequate to covering all costs for their members. However, according to their own calculations, based on accepting publishers' quoted estimates of between \euro{1000} and  \euro{2000} per article as the appropriate average cost,  this would not decrease the overall fees to participating members, but rather double them from an estimated current total of  \euro{5m} per year to \euro{10m}, assuming a base of 500 participating institutions, 
   
   Within an increasingly efficient, web-based, low cost publishing environment, this approach could nevertheless help bring about a more realistic correspondence between subscription and publication charges and actual production costs. In this setting smaller, low production cost OA publishers could possibly pose a competitive threat to publishing giants like Elsevier or Springer if the latter are unable to adapt their pricing strategies.  But even if SCOAP${}^3$ succeeds in providing an effective alternative that helps reverse trends towards excessive journal charges within the domain of particle physics, it is not clear whether this could extend to other areas.  There are few other domains where common interests and shared facilities are so coherently defined, or where the number of journals in which most published research appears is so small.  Other alternatives and models for scientific publishing must therefore also be considered, and efforts made to bring market forces to bear in the interests of the community.

%%%%%%%%%%%%%%%%%  Section 4  %%%%%%%%%%%%%
\section{Conclusions}
    
   Amongst current models, the shortcomings of the ``Author pays" approach to OA scientific publishing are clear: it further taxes scarce research funds by transferring the cost burden to authors, excludes those who are unwilling or unable to pay such charges, and places the implementation of peer review standards in a competitive relation with the financial viability of the journal. The research community provides not only the published material, but also the refereeing services that, with distribution, form the main ``value-added" that journals offer. It is therefore up to its members to make the choices, and exert the necessary pressure to assure that they are beneficiaries of transformations in scientific publishing that are occurring as a result of evolution in technology and consumer reaction against inflated subscription prices.
   
    Meanwhile, at least in most branches of physics, mathematics, computer science and some other domains, the benefits of Open Access will continue to be adequately provided by widely used repositories such as arXiv. Other fields seeking to develop effective vehicles for Open Access could do well to first consider such an approach,  where preprint/postprint repositories have long been seen to coexist constructively with refereed journals, providing in a complementary fashion for rapid dissemination, universal access,  assurance of quality standards and long-term preservation of the results of scientific research.
    
\end{document}